\documentclass[12pt]{article} 
\usepackage{amsmath} 
 
\begin{document} 
 
\def \be {\begin{equation}} 
\def \ee {\end{equation}} 
\def \bea {\begin{eqnarray}} 
\def \eea {\end{eqnarray}} 
\def \bse {\begin{subequations}} 
\def \ese {\end{subequations}} 
\def \bde {\begin{description}} 
\def \ede {\end{description}} 
\def \nn {\nonumber} 
\def \spa {$\;$} 
\def \la {\langle} 
\def \ra {\rangle} 
\def \R {{\bf R}} 
\def \C {{\bf C}} 
\def \Z {{\bf Z}} 
\def \del {\partial} 
\def \dels {\partial\kern-.5em / \kern.5em} 
\def \As {{A\kern-.5em / \kern.5em}} 
\def \Ds {D\kern-.7em / \kern.5em} 
\def \Psib {{\bar \Psi}} 
 
\def \rh {\kappa} 
\def \a {\alpha} 
\def \b {\beta} 
\def \dag {\dagger} 
\def \g {\gamma} 
\def \G {\Gamma} 
\def \d {\delta} 
\def \eps {\epsilon} 
\def \m {\mu} 
\def \n {\nu} 
\def \k {\kappa} 
\def \lam {\lambda} 
\def \Lam {\Lambda} 
\def \s {\sigma} 
\def \r {\rho} 
\def \om {\omega} 
\def \Om {\Omega} 
\def \one {{\bf 1}} 
\def \th {\theta} 
\def \Th {\Theta} 
\def \Chi {\chi} 
\def \t {\tau} 
\def \ve {\varepsilon} 
\def \II {I\hspace{-.1em}I\hspace{.1em}} 
\def \IIA {\mbox{\II A\hspace{.2em}}} 
\def \IIB {\mbox{\II B\hspace{.2em}}} 
 
\newcommand{\ho}[1]{$\, ^{#1}$} 
\newcommand{\hoch}[1]{$\, ^{#1}$} 
 
 
\def \H {{\cal H}} 
\def \MSUSY {M_{\mbox{\tiny SUSY}}}. 
\def \S {{\cal S}} 
\def \ha {\hat{a}} 
\def \u {u} 
\def \z {\om} 
\def \bigdot {\mbox{\Large .}}
 
 
\begin{center} 
\hfill hep-th/0312288\\ 
\vskip .5in 
 
\textbf{\large 
Large N Cosmology 
} 
 
\vskip .5in 
{\large
Robert Brandenberger $^a$,
Pei-Ming Ho $^b\, ^c$
\footnote{on leave from National Taiwan University},
Hsien-chung Kao $^d$}
\vskip 15pt 
 
{\small \em $^a$ Department of Physics, 
Brown University, Providence, RI.}\\  
{\small \em $^b$ Department of Physics, 
National Taiwan University, Taipei, Taiwan, R.O.C.}\\ 
{\small \em $^c$ Department of Physics, 
Harvard University, Cambridge, MA.}\\ 
{\small \em $^d$ Department of Physics, 
National Taiwan Normal University, Taipei, Taiwan, R.O.C.}\\ 
\vskip .2in 
\sffamily{ 
rhb@het.brown.edu \\
pmho@phys.ntu.edu.tw \\ 
hckao@phy.ntnu.edu.tw}
 
\vspace{60pt} 
\end{center} 
\begin{abstract}

We motivate inflationary scenarios with many scalar fields, 
and give a complete formulation of adiabatic and entropy perturbations. 
We find that if the potential is very flat, 
or if the theory has a $SO(N)$ symmetry, 
the calculation of the fluctuation spectrum can be carried out 
in terms of merely two variables without any further assumption. 
We do not have to assume slow roll or 
$SO(N)$ invariance for the background fields. 
We give some examples to show that, 
even if the slow roll assumption holds, 
the spectrum of fluctuations can be quite different from 
the case when there is a single inflaton.

\end{abstract} 
 
\setcounter{footnote}{0} 
\newpage 
 
\section{Introduction} 
 
As phenomenological scenarios of the early Universe, 
inflationary models constructed with a single inflaton 
are popular for their simplicity and consistency 
with existing observations. 
However, for a coherent understanding of the Universe, 
cosmological models should be embedded in fundamental particle theories. 
As we will argue now, 
models with many inflatons may be more natural than 
those with a single inflaton. 
 
It is often natural to have a large number of fields 
in various models for the early Universe. 
In Kaluza-Klein theories or string theory, 
it is very common to have a large number of moduli fields. 
For some \cite{Susskind},
the enormous moduli space is
the most important feature of string theory.
For example, in the brane world scenario, 
there is a scalar field associated with each extra dimension, 
corresponding to brane fluctuations in that direction. 
In addition, for $n$ D-branes on top of each other, 
each of these scalar fields is promoted to a $n\times n$ matrix. 
There are $6\cdot 5^2 = 150$ independent real scalar fields 
for 5 D3-branes in 10 dimensions, for instance. 
As another example, for the tachyon model of inflation \cite{Gibbons}, 
it was pointed out that it is hard to be consistent 
with observations \cite{KL}, 
unless there is a large number of coincident branes \cite{PCZZ}. 
In fact, merely the 10 dimensional metric already gives 
6 vectors and 15 scalars in addition to the metric in 4 dimensions. 
Recently, the possibility of inflation in pure gravity 
due to compactifications on time-varying hyperbolic spaces 
has attracted some attention and is being explored \cite{hyper}. 
The scalars coming from the higher dimensional metric 
play the role of inflatons in these models. 
Based on the above examples, it may be quite 
unnatural to have a model of inflation with a single inflaton field. 
In fact, even the standard model has 3 or 4 real components
from the Higgs doublet;
and there are generically a large number of scalar fields
in any grand unified theory (GUT).
\footnote{
In $SU(5)$ GUT, there is a Higgs field in the adjoint representation
with $24$ components,
and another in the fundamental representation with $5$ components.
There are a lot more Higgs fields for $SO(10)$ GUT's.}

Another motivation for multiple inflatons comes 
from the observation that in order to have a scale invariant 
spectrum of density perturbations, 
we need the slow roll condition 
$\left|\frac{\dot{H}}{H^2}\right|\ll 1$ to be satisfied. 
This implies that we need the potential energy of the inflatons 
to dominate over the kinetic energy. 
A naive scaling of the potential and kinetic energy terms 
with the number $N$ of inflatons 
indicates that this condition is automatically satisfied in the 
large $N$ limit, as we will show in Sec.\ref{LargeN}. 
\footnote{However, this scaling does not apply to assisted inflation.} 

Inflation driven by multiple scalar fields is not a new idea.
One of the examples which has been considered is the 
so-called ``assisted inflation'' scenario \cite{Salopek1,LMS}, 
where exponential potentials are analyzed, 
and it was found that inflation is more easily achieved 
when there are a large number of scalar fields.
As another example, density perturbations of many scalar fields which interact with
one another only through gravity (but can have arbitrary self-interacting potentials)
were studied in \cite{PS}.
General analytic formulae for density perturbations in
the multi-field inflation scenario have already been derived \cite{SSS,FB,GWBM}.

However, as we will see,
it is in general a lot more complicated to deal with more than one field. 
In the case of a single inflaton, the slow roll condition 
implies that the slope of the potential is small. 
But this may not be true when there are more scalar fields,
as will be explained  in Sec.\ref{SR}.
Furthermore, for $N$ scalar fields,
one needs to solve $N$ coupled 2nd order differential equations
for the background fields,
and $N$ additional coupled equations for the perturbations of each field.
For a large number $N$ this can be a formidable task even with a computer.

Despite the complexity of generic multiple inflaton models, 
the system can be greatly simplified in \underline{special case}s.
In \cite{Salopek1} it was found that the background evolution only
depends on a single variable if the scalars are free fields
(and the inflation is driven by the cosmological constant).
Later it was shown \cite{MW} that,
for exponential potential, many scalar fields can be assembled
into an adiabatic and an entropy mode by making a suitable rotation
in field space,
and the same result was obtained for polynomial potentials $\sum_i V(\phi_i)$ in \cite{KO},
where each field has a potential of the same functional form $V$.
In this paper, we extend the analysis to the non-trivial case
when the inflatons observe an $SO(N)$ global symmetry.
We find that only a single variable is needed to describe
the evolution of the background geometry, and, more remarkably,
only two equations for two independent variables are needed 
for the purpose of determining adiabatic and entropy perturbations.
Note that, although we assume $SO(N)$ symmetry for the interaction,
the background field configuration does not have to be $SO(N)$ invariant. 

More specifically, we will focus on the case 
when the scalar fields are in the fundamental representation 
of the $SO(N)$ symmetry.
We expect that the same technique can be applied to more general cases 
with different $SO(N)$ representations, 
or when the symmetry group is different. 
In those cases the number of variables needed for 
a complete description of the perturbations may be larger than two. 
Our experience with the simplest case 
will be valuable for those generalizations. 

Examples of scalar fields in the fundamental representation
are not hard to find.
The Higgs doublet in the standard model has four real components,
which transform as a fundamental representation under an $SO(4)$ symmetry,
for which the Higgs potential is invariant.
The Higgs potentials in grand unified theories
do not observe any $SO(N)$ symmetry in general.
But if we consider the weak field approximation,
and keep only the quadratic terms in the potential,
$SO(N)$ symmetry often emerges.
For the fundamental and adjoint representations of $SU(5)$, for instance,
the quadratic terms in the Higgs potential in $SU(5)$ GUT,
$\mbox{Tr}(H^{\dag}H)$ and $\mbox{Tr}(\Sigma^2)$,
have $SO(10)$ and $SO(24)$ symmetries, respectively.
For the brane-world scenario,
there are scalar fields in the fundamental representation of $SO(N)$
if there are extra $N$ (flat) dimensions.

Imposing a symmetry on the model not only simplifies 
the formulation and makes calculations accessible, 
it also helps to render the model consistent with 
observational constraints on the magnitude of the entropy perturbation
mode. Without the symmetry, 
each matter field gains different amounts of fluctuations 
according to its coupling to each inflaton. 
The predicted CMB perturbations will obtain contributions 
not only from the adiabatic mode of the metric perturbations, but also 
from the perturbations of each matter field that interacts with photons. 
Observational data shows that the adiabatic mode dominates, 
and thus we have to avoid excessive isocurvature perturbations. 
However, if the coupling of each matter field to 
the inflatons is $SO(N)$ invariant, 
only $SO(N)$ invariant operators can contribute to CMB fluctuations.
Hence, we argue that inflationary
models with multiple inflatons and with a large symmetry group 
are the most promising models to study. 
Similar conclusion was also reached in \cite{HHR}.

This paper is organized as follows.
After reviewing the general formulation for 
multiple scalar fields coupled to gravity, 
we give the evolution equations for the background in Sec.\ref{LargeN}, 
and formulae for adiabatic and entropy perturbations in Sec.\ref{pert}.
We studied perturbations for the following two circumstances:
(1) theories with very flat potentials (Sec.\ref{SP}), 
(2) theories with $SO(N)$ symmetry  (Sec.\ref{SON}). 
In the first case, 
if we in addition to the condition on the potential assume the 
slow roll condition, 
we obtain the same scale invariant spectrum of metric perturbations 
as the case of a single inflaton. 
This result strengthens the case for the robustness of the connection 
between slow-roll inflation and having a scale invariant spectrum
of fluctuations. 
 
In the second case, in which the fields have a $SO(N)$ symmetry,
we find that even if the background configuration is arbitrary, 
we still only need two variables to describe the adiabatic and
entropy fluctuations. This is true because 
the coupling of the metric to the scalars is $SO(N)$ invariant. 

Throughout this paper we will adopt the convention $8\pi G = 1$.

\section{Background Equations and Large N Limit} \label{LargeN} 
 
In this section we
review
the evolution 
of homogeneous and isotropic configurations of 
$N$ scalar fields coupled to Einstein gravity. 
 
For $N$ scalar fields with the Lagrangian density 
\be 
{\cal L} =  
\sum_I \frac{1}{2}g^{\mu\nu}\del_{\mu}\phi_I\del_{\nu}\phi_I 
- V(\phi) 
\ee 
in a spatially flat Robertson-Walker Universe 
with the metric 
\be 
ds^2 = dt^2-a^2(t) dx^2 = a^2 (d\eta^2 - dx^2), 
\ee 
(the variable $t$ denotes physical time whereas $\eta$ is
conformal time)
the equations of motion in the homogeneous background are 
(with an over-dot denoting the derivative with respect to $t$)
\bea  
&\;& \r = K+V = 3 H^2,  \label{EOMa1}\\ 
&\;& p = K-V = -2\dot{H} - 3H^2,  \label{EOMa2}\\ 
&\;& \ddot{\phi}_I+3H\dot{\phi}_I+V_I = 0. \label{EOMphi} 
\eea 
Here, $H \equiv \frac{\dot{a}}{a}$ is the Hubble rate, 
$K \equiv \sum_{I} \frac{1}{2}\dot{\phi}_I^2$ and $V$ are the kinetic and
potential energy densities, respectively, and  
$V_I \equiv \frac{\del V}{\del \phi_I}$. 

{F}rom (\ref{EOMa1}) and (\ref{EOMa2}), 
the time evolution of $K$ and $V$ 
is completely known if the scale factor $a(t)$ is given 
\bea  
&\;& K = -\dot{H}, \label{K} \\ 
&\;& V = \dot{H}+3H^2. \label{V} 
\eea 
As is well-known, not all of the equations 
(\ref{EOMa1}), (\ref{EOMa2}), and (\ref{EOMphi}) 
are independent. 
Indeed, by multiplying (\ref{EOMphi}) by $\dot{\phi}_I$ 
and summing over $I$ we see that eq. (\ref{EOMa2}), 
the spatial part of the Einstein equation, 
can be derived upon using (\ref{EOMa1}).  
 
Conversely, $H$ and $\dot{H}$ are determined 
by $K$ and $V$ via (\ref{K}) and (\ref{V}). 
In particular, the ratio $\dot{H}/H^2$ is 
determined by the ratio $K/V$. 
We have smaller $\dot{H}/H^2$ for smaller $K/V$. 
In the inflationary scenario, 
$\dot{H}/H^2$ has to be small in order to have 
a scale invariant spectrum of density perturbation, 
implying that the potential energy must dominate over 
the kinetic energy. 
Below, we will see that potential energy domination
arises naturally in the large $N$ limit, as was
already realized in \cite{Mazenko} (see also \cite{rey}). 
 
For a system of $N$ scalar fields coupled to gravity, 
the kinetic term is proportional to $N$. 
A generic potential term of the form 
$\sum_{I_1\cdots I_n}\lam_{I_1\cdots I_n} \phi_{I_1}\cdots\phi_{I_n}$ 
is of order $N^n$ 
because it involves $n$ consecutive sums over $N$ fields. 
For $n > 1$, which holds unless the theory is free, 
the potential energy dominates over the kinetic energy. 
Inflation is thus a natural consequence of the large $N$ limit. 
 
The argument above is based on the assumption that the field
values $\phi_I$ and the coupling constants $\lam_{I_1\cdots I_n}$ 
do not scale with $N$. The underlying philosophy is that, in a 
fundamental theory, all coupling constants are expected to be of order $1$ 
in fundamental units. In string theory, for instance, 
the couplings among the fields are determined by the string coupling $g_s$, 
which is set by the expectation value of the dilaton field. 
If we include the dilaton as one of the scalar fields, then the natural 
assumption is that all couplings are of order $1$. 
Note that one aspect of our assumption says that 
the initial values for all fields should be of order $1$. This
assumption is related to the types of arguments 
used to motivate the initial conditions for chaotic inflation \cite{Linde}
(in models of chaotic inflation, the stages of inflation important
for comparison with observations - the last 60 e-foldings of inflation -
take place when the field values are of the order $1$ in Planck units). 
 
Notice that this large $N$ limit is different from that of 't Hooft, 
where the coupling constants are chosen to scale with $N$ 
in a particular way such that quantum effects 
have a well defined limit when the classical background is set to zero. 
For the $\phi^4$ theory, for example, the coupling constant $\lam$ scales 
like $1/N$. The target problem of 't Hooft's large $N$ limit 
is very different from our consideration of the 
dynamics of classical background fields. 
 
Having a large number of fields means that, in general, the computations 
which have to be done in cosmology are more complicated. 
For instance, we will see below that, to calculate the metric perturbations, 
we need to solve $N$ coupled differential equations. 
We will consider two circumstances in which the system of a 
large number of field fluctuations 
can be solved in terms of only two fields. 
 
\section{Adiabatic and Entropy Perturbations} \label{pert} 

In this section we give the general formulation 
for the generation and evolution of perturbations 
for models with multiple inflatons.
Although equivalent formulae were given before in \cite{SSS,FB,GWBM},
here we present them for completeness and also take this opportunity
to introduce the variables $\kappa$ and $\mu$ that will be useful later.

\subsection{Adiabatic Perturbation} 
 
Fluctuations of the scalar fields $\delta \phi_I$ 
induce fluctuations of the metric. 
In the longitudinal gauge (see e.g. \cite{MFB} for a survey of
the theory of cosmological fluctuations), 
the metric describing scalar metric degrees of freedom is of the form 
\be 
ds^2 = (1-2\Phi)dt^2-a^2(t)(1+2\Psi)dx^2 \, , 
\ee 
where the two functions $\Phi$ and $\Psi$ representing the
fluctuations are functions of space and time.
For matter consisting of $N$ scalar fields, there are to linear
order no non-vanishing off-diagonal space-space components of the
energy-momentum tensor. Hence, it follows from the off-diagonal
space-space components of the Einstein equations that $\Phi=\Psi$. 
Vector metric perturbations decay in an expanding Universe and
will thus not be considered here. Tensor metric perturbations
(gravitational waves) do not couple to matter and their evolution 
will hence be as in single-field inflation models.
 
The Einstein equations and the equations of motion 
for the scalar fields at lowest order are given by 
(\ref{EOMa1}) and (\ref{EOMphi}). 
The independent parts of the linear fluctuation equations 
take the form (see e.g. \cite{FB}) 
\bea 
&-3H\dot{\Phi}-\frac{k^2}{a^2}\Phi = 
\sum_I\frac{1}{2}(\dot{\phi}_I\delta\dot{\phi}_I+V_I\delta\phi_I) 
+V(\phi)\Phi, \label{dPhi1} \\ 
&\dot{\Phi}+H\Phi = 
\sum_I\frac{1}{2}\dot{\phi}_I\delta\phi_I, \label{dPhi2} \\ 
&\delta\ddot{\phi}_I+3H\delta\dot{\phi}_I 
+\sum_J\left(\frac{k^2}{a^2} \delta_{IJ} + V_{IJ} 
\right)\delta\phi_J = 
4\dot{\Phi}\dot{\phi}_I - 2V_I \Phi. \label{ddphi} 
\eea 
In linear perturbation theory, each Fourier mode evolves
independently. Hence, it is easier to solve the equations
in Fourier space (i.e. after having performed a Fourier transformation 
from comoving spatial coordinates $x$ to comoving wave numbers $k$). 
Note that the variables $\Phi$ and $\phi_I$ have a $k$-dependence 
which is not manifest in our notation. 
 
The Sasaki-Mukhanov \cite{SM} variables 
for the $N$ scalar fields are 
\be 
Q_I \equiv \delta\phi_I + \frac{\dot{\phi}_I}{H} \Phi. 
\ee 
In terms of these variables, the above equations become (see e.g. \cite{FB}) 
\bea  
&\;& -\frac{k^2}{a^2}\Phi - (3H^2+\dot{H})\left\{\left(\frac{\Phi}{H}\right)^{\bigdot}  
+\Phi \right\} = 
\sum_I\frac{1}{2}(\dot{\phi}_I \dot{Q}_I+V_I Q_I), \label{EOMQ1}\\ 
&\;& H \left\{\left(\frac{\Phi}{H}\right)^{\bigdot}  
+\Phi \right\} = 
\sum_I\frac{1}{2}\dot{\phi}_I Q_I, \label{EOMQ2}\\ 
&\;& \ddot{Q}_I + 3H \dot{Q}_I + \frac{k^2}{a^2} Q_I 
+ \sum_J M_{IJ} Q_J = 0. \label{EOMQ} 
\eea 
Here 
\bea 
M_{IJ} &=& V_{IJ} - \frac{1}{a^3} \left( 
\frac{a^3}{H}\dot{\phi}_I\dot{\phi}_J 
\right)^{\bigdot} \nn\\ 
&=& V_{IJ}+\frac{1}{H}\left[ 
\left(\frac{\dot{H}+3H^2}{H} 
\right)\dot{\phi}_I\dot{\phi}_J 
+V_I\dot{\phi}_J+V_J\dot{\phi}_I 
\right]. \label{MIJ} 
\eea 
The comoving curvature perturbation is defined by 
\be \label{zeta} 
\zeta \equiv  
= -\frac{H^2}{\dot{H}}\left\{\left(\frac{\Phi}{H}\right)^{\bigdot} 
+\Phi \right\} \, .
\ee 
As shown in \cite{Taruya}, in the multi-field case $\zeta$ becomes
\be
\zeta = \frac{H}{2}\frac{\sum_I \dot{\phi}_I Q_I}{K} \, .
\ee
The variable $\zeta$ is related to the intrinsic three-curvature
on the constant energy density surface $\Sigma$ via 
\be 
 ^{^{(3)}}R = \frac{-4k^2}{a^2}\Psi_{\Sigma},  
\ee 
where $\Psi_{\Sigma}= \zeta - k^2 \Psi/(3a^2 \dot{H})$.  
We can use either $\zeta$ or equivalently $\Phi$ as 
a measure of the curvature perturbation, which represents the adiabatic 
mode of the metric perturbations. 
 
We now introduce the two gauge-invariant variables 
\be 
\rh = \sum_I \dot{\phi}_I Q_I, \quad 
\mu = \sum_I V_I Q_I . \label{rhomu} 
\ee 
The curvature fluctuation can be written in terms of these two variables as 
\be \label{zetaPhi} 
\zeta = - \frac{H\kappa}{2\dot{H}}, \quad 
\Phi = -\frac{a^2}{2k^2}\left\{ 
\dot{\rh}+\left(6H+\frac{\dot{H}}{H}\right)\rh+2\mu \right\}. 
\ee 
Conversely, we can also express $\rh$ and $\mu$ in terms of $\Phi$: 
\bse 
\label{rho_mu} 
\bea  
&\;& \rh= 2H \left\{\left(\frac{\Phi}{H}\right)^{\bigdot} +\Phi \right\}, \\ 
&\;& \mu =-\frac{a^2}{k^2}\Phi - (6H^2-4\dot{H} -\frac{\ddot{H}}{H})\Phi  
- 7H \dot{\Phi} - \ddot{\Phi}.  
\eea 
\ese 

In the following subsection we will see that also the entropy fluctuation
can be written in terms of the two variables $\mu$ and $\kappa$. 
 
\subsection{Entropy Perturbation} 
 
A dimensionless gauge-invariant 
definition of the total entropy perturbation is (see e.g. \cite{WMLL})
\be 
\S = H\left(\frac{\d p}{\dot{p}} 
- \frac{\d \r}{\dot{\r}}\right). 
\ee 
For $N$ scalar fields, this yields
\be 
\S = \frac{2(\dot{V}+3H\sum_I \dot{\phi}_I^2)\d V 
+ 2\dot{V}\sum_I\dot{\phi}_I(\d\dot{\phi}_I-\dot{\phi}_I\Phi)} 
{3(2\dot{V}+3H\sum_I \dot{\phi}_I^2)\sum_J \dot{\phi}_J^2}. 
\ee 
In terms of $\rh$ and $\mu$, we have \cite{GWBM}
\be 
\S = - \frac{2(\ddot{H}+3H\dot{H})\mu 
+(\ddot{H}+6H\dot{H})\left[\dot{\rh} 
+\frac{\dot{H}+3H^2}{H}\rh\right]} 
{6\dot{H}(\ddot{H}+3H\dot{H})}. 
\ee 
We see that both adiabatic and entropy 
perturbations can be expressed in terms of 
the two variables $\rh$ and $\mu$.   
 
On the other hand, if we express the entropy perturbation in terms of 
$\Phi$, we have 
\be \label{entropy}
\S = \frac{k^2}{3a^2 \dot{H}}\Phi - \frac{H^2}{3H \dot{H} + \ddot{H}} \left\{\left(\frac{\Phi}{H}\right)^{\bigdot} +\Phi \right\}. 
\ee 
In the long wave length limit, it turns out that \cite{GBW,FB,WMLL,GWBM} 
\be \label{zetaS} 
\dot{\zeta} \simeq \frac{\dot{p}}{\r+p}\S. 
\ee 
Assuming the slow roll condition ($\dot{H} \ll H^2$), the above
becomes 
\be \label{zetaS2} 
\dot{\zeta} \simeq 
3H\left(\frac{\dot{p}}{\dot{\r}}-\frac{\delta p}{\delta\r}\right)\zeta. 
\ee 
In the case of a single inflaton, 
$\S$ vanishes and curvature perturbations $\zeta$ are frozen 
outside the Hubble radius. 
 
In the case of multiple fields, the entropy perturbation is in 
general not zero, and it can be important in determining 
the magnitude of adiabatic perturbation. For example, as shown
in \cite{FB} and \cite{BassettVin}, entropy fluctuations can
seed exponential growth of super-Hubble curvature fluctuations during
inflationary reheating. In our case, when 
the inflatons are slowly rolling down the potential hill, 
we have $\dot{p}>0$ and $(\r+p)>0$. Since ${\dot H} < 0$, 
Equations (\ref{entropy}) and (\ref{zetaS}) imply that 
the amplitude of super-Hubble-scale curvature perturbations will 
increase due to the coupling to the isocurvature mode.
 
\subsection{Generation of Quantum Fluctuations} 
 
Since the inflationary expansion redshifts all classical
perturbations pre-existing before inflation, it leaves
behind a quantum vacuum. In the context
of inflationary cosmology, quantum vacuum fluctuations are
believed to be the origin of the fluctuations we observe
today. In quantizing the cosmological perturbations, it is
important to identify the variables in terms of which
the action for fluctuations takes on canonical form, i.e.
in terms of which the fluctuation fields have the usual
kinetic term. As reviewed in \cite{MFB}, the canonical
variables $v_I$ are related to the $Q_I$ variables via
$v_I = a(t) Q_I$. Hence, if we perform standard canonical
quantization, and expand the canonical fields into
creation and annihilation operators, then the expansion
of the fields $Q_I$ in terms of these operators takes the form 
(at the initial time $t_0$) 
\bea \label{t0} 
Q_{I}(t_0)&=&\frac{1}{\sqrt{2k}a(t_0)} 
\left(\ha_{I}(t_0)+\ha_{I}^{\dag}(t_0)\right), \\ 
P_{I}(t_0)&=&-ia(t_0)\sqrt{\frac{k}{2}} 
\left(\ha_{I}(t_0)-\ha_{I}^{\dag}(t_0)\right), 
\eea 
where the $P_I$ are the momenta conjugate to $Q_I$ 
\be \label{momP} 
P_I(t) = a^2(t) \dot{Q}_I(t) + a^2(t) H Q_I(t). 
\ee

Since the evolution of $Q_I$ (\ref{EOMQ}) 
satisfies linear differential equations, 
$Q_I$ and $P_I$ at a later time $t$ must depend 
on the initial data linearly, 
\bea 
Q_{I}(t)&=&\frac{1}{\sqrt{2k}a(t)} 
\sum_J \left(q_{IJ}(t) \ha_{J}(t_0)+\bar{q}_{IJ}(t) \ha_{J}^{\dag}(t_0)\right), \\ 
P_{I}(t)&=&-ia(t)\sqrt{\frac{k}{2}} 
\sum_J \left(p_{IJ}(t) \ha_{J}(t_0)-\bar{p}_{IJ}(t) \ha_{J}^{\dag}(t_0)\right), 
\eea 
where the functions
$q_{IJ}(t)/a(t)$ satisfy (\ref{EOMQ}) 
for all $J$ and 
\be \label{pq} 
p_{IJ}=i\frac{\dot{q_{IJ}}}{k}. 
\ee

Due to (\ref{t0}), the initial conditions are 
\be \label{init1} 
q_{IJ}(t_0) = p_{IJ}(t_0) = \d_{IJ}. 
\ee 
To solve the differential equation for $q_{IJk}(t)$, 
we need the initial values for both $q_{IJk}(t_0)$ and $\dot{q}_{IJk}(t_0)$. 
The latter are (according to (\ref{pq})) given by 
\be \label{init2} 
\dot{q}_{IJ}(t_0) = - i k \d_{IJ} . 
\ee  
 
The two point correlation function at time $t$ 
for the vacuum state defined at $t_0$, 
denoted $|0\ra_0$, is 
\be 
{}_0\la 0| Q_{I}(t) Q_{J}(t) |0\ra_0 = 
\frac{1}{2ka^2(t)}|\sum_K q_{IK}(t) q^{\dag}_{KJ}(t)|\d^{(3)}(k-k'). \label{2pt} 
\ee 
(On the left hand side, 
we omitted the $k$-dependence of $Q_I$ and $Q_J$, 
which should be denoted $Q_{Ik}$ and $Q_{Jk'}$.) 
 
As we have seen in the previous section, 
the quantities $\zeta$, $\Phi$ and $\S$, 
which we are interested in, 
can be expressed in terms of $\rh$ and $\mu$ (\ref{rhomu}). 
Consequently, their correlation functions can also be expressed  
in terms of those of $\rh$, $\mu$ and their derivatives.  
 
In general, this calculation involves solving 
the equations of motion for each of the $N$ fields $Q_I$. 
However, we will see in the following sections that, 
in certain cases, the evolution equations of $\rh$ and $\mu$ form   
a set of coupled differential equation by themselves.   
As a result, they can be completely determined 
without having to know the evolution of individual components $Q_I$. 
%
 
\section{Slow Roll} \label{SR} 
  
The ``slow roll'' conditions are usually defined as 
\be \label{slowroll} 
|\dot{H}| \ll H^2, \quad |\ddot{H}| \ll |\dot{H}H|. 
\ee 
It then follows from (\ref{K}), (\ref{V}) that 
\be 
V \gg K, \quad \dot{V} \gg \dot{K}. 
\ee 
 
For the case of a single inflaton, 
the slow roll conditions imply that $|\frac{dV}{d\phi}| \ll V$, since  
\be 
|\frac{dV}{d\phi}| = \frac{|\dot{V}|}{|\dot{\phi}|} 
     \simeq \frac{6|\dot{H}|H}{\sqrt{2K}} 
     \simeq 3\sqrt{2|\dot{H}|}H \ll \sqrt{2} V. 
\ee 
Similarly one can easily show that $|\frac{d^2 V}{d\phi^2}| \ll V$. 
In this case, the slow roll conditions also imply that the second
time derivative of the scalar field is negligible. We will now
show that this last conclusion 
does not hold if there are many inflaton fields
(in which case the use of the term ``slow rolling'' is in fact
somewhat misleading).
 
In the case of a large number of inflatons, 
we still need the vector $\dot{\phi}_I$ to be small, 
so that $K \ll V$. 
However, we can arrange $\ddot{\phi}_I$, 
viewed as an $N$-vector, to be roughly perpendicular  
to $\dot{\phi}_I$ such that $\ddot{H}=-\dot{\phi}_I\ddot{\phi}_I$  
is nearly zero.  This can occur even if $\ddot{\phi}_I$ is not small. 
(Note that $\dot{\phi}_I$ can not be directly related to $V_I$ 
if $\ddot{\phi}_I$ is not small.) 
In addition, since $\dot{\phi}_I$ may be 
vanishingly small at a certain moment, 
$V_I = \dot{V}/\dot{\phi}_I$ can be 
very large compared to $V$. 
Therefore, it may not be a good approximation to ignore $M_{IJ}$ (\ref{MIJ})
even if the slow roll conditions are satisfied, and 
the calculation for fluctuations will in general be much more complicated 
than in the case of a single inflaton. 
 
In the following we consider two situations in which 
the theory of perturbations is greatly simplified, and in 
which we will not have to solve $N$ coupled equations for each $Q_I$, 
but only have two coupled equations for $\rh$ and $\mu$.

\section{Smooth Potential} \label{SP} 
 
Following the previous section, we see that the ``slow roll" conditions 
are not enough to guarantee the scale invariance of the spectrum 
of fluctuations if there are more than one inflatons, the reason
being that the $M_{IJ}$ terms in  Equation (\ref{MIJ}) are
not in general negligible. 
In this section we consider the case when the potential 
is very flat and show that adiabatic and entropy perturbations 
can be determined by the two variables $\rh$ and $\mu$ alone. 
The scale invariance of the spectrum will result 
if we assume that the slow roll conditions hold. 
 
Suppose that the potential is very flat, more precisely, 
\bea 
&V_{IJ} \ll \dot{\phi}_I\dot{\phi}_J, \label{VIJ} \\ 
&V_{IJK} \ll \dot{\phi}_I\dot{\phi}_J\dot{\phi}_K \label{VIJK} 
\eea 
in Planck units. 
It follows from (\ref{VIJ}) that 
\be 
M_{IJ}\simeq\frac{1}{H}\left( 
\frac{\dot{H}+3H^2}{H}\dot{\phi}_I\dot{\phi}_J 
+V_I\dot{\phi}_J+V_J\dot{\phi}_I\right). 
\ee 
It is straightforward to derive 
the equations of motion for $\rh$ and $\mu$ 
\bea 
&\ddot{\rh}+9H\dot{\rh}+ 
\left(\frac{k^2}{a^2}+18H^2+3\dot{H} 
-2\frac{\dot{H}^2}{H^2}+\frac{\ddot{H}}{H}\right)\rh+ 
2\dot{\mu}+\frac{6H^2-2\dot{H}}{H}\mu \simeq 0, \\ 
&\ddot{\mu}+3H\dot{\mu}+ 
\left(\frac{k^2}{a^2}+6\dot{H} 
+\frac{\ddot{H}}{H}\right)\mu 
-\left(\frac{H^{(3)}}{H}+6\ddot{H} 
-\frac{\dot{H}\ddot{H}}{H^2}\right)\rh \simeq 0. 
\eea 
In deriving the above we used 
(\ref{V}) and 
\be 
V_I V_I \simeq -(\ddot{V}+3H\dot{V}), 
\ee 
which in turn can be derived from Equation (\ref{EOMphi}) 
by multiplying it with $V_I$ and making use of Equation (\ref{VIJ}). 
 
If we also assume the slow roll conditions (\ref{slowroll}), 
the equations are simplified to 
\bea 
&\hat{\rh}''+(k^2-2\H^2)\hat{\rh}+2a^4(\hat{\mu}'+2\H\hat{\mu}) = 0, 
\label{rh1}\\ 
&\hat{\mu}''+(k^2-2\H^2)\hat{\mu} = 0,  \label{mu1} 
\eea 
where 
\be \label{hat} 
\hat{\rh} = a^4 \rh, \quad \hat{\mu} = a\mu, 
\quad \H = \frac{a'}{a}, 
\ee 
and the primes refer to derivatives with respect to conformal time 
\be 
d\eta = \frac{dt}{a}. 
\ee 
The slow roll conditions imply 
\be 
\H' \simeq \H^2 \simeq \frac{a''}{2a}. 
\ee 

Equation (\ref{mu1}) is the key equation needed to demonstrate
that a scale-invariant spectrum results, as we now show. 
A fluctuation mode with wave number $k$ crosses 
the Hubble radius at a time $\eta_k$ when $k\simeq\H$, 
or equivalently when
\be \label{etak} 
H = \frac{k}{a(\eta_k)}. 
\ee 
For $\eta \ll \eta_k$, so that $k \gg \H$, 
we see from Equations (\ref{rh1}) and (\ref{mu1}) that 
the fluctuation is in the oscillatory phase. 
The amplitude of the canonical fluctuation variable remains the same. 
Hence, the amplitude of the fluctuations when they cross the Hubble
radius is the vacuum amplitude, and as in the single inflaton case
this implies that the spectrum will be scale invariant.

To see this explicitly, consider that
when $\eta \gg \eta_k$, so that $k \ll \H$, 
the fluctuation $\hat{\rh}$ grows like $a$. 
The approximate solution of the equations is
\be \label{scaling}
\hat{\mu} = \hat{\mu}_0\frac{a(\eta)}{a(\eta_k)}, \quad 
\hat{\rh} = \hat{\rh}_0\frac{a(\eta)}{a(\eta_k)} 
-\frac{\hat{\mu}_0}{3a(\eta_k)}\frac{a^5(\eta)}{\H}, 
\ee 
where $\hat{\rh}_0$ and $\hat{\mu}_0$ are the initial values of 
$\hat{\rh}$ and $\hat{\mu}$ 
prescribed at a moment $\eta_0 < \eta_k$. 
To find the dependence on $k$ of the two point correlation function of 
$\zeta$ (or $\Phi$), we use (\ref{zetaPhi}) and (\ref{2pt}). 
Since the dominant mode of $\hat{\rh}$ scales as $a(t)^4$ (from
Equation (\ref{scaling}), $\rh$ is independent of time, and hence 
\be 
{}_0\la 0|\zeta_k \zeta_k|0\ra_0 \propto 
\frac{1}{2k a^2(\eta_k)} \propto \frac{1}{k^3}, 
\ee 
where we employed (\ref{etak}). 
This corresponds to a scale invariant spectrum. 
 
Let us remark here that 
a scale invariant spectrum at the Hubble radius during inflation 
does not always imply the same for the observed spectrum 
of density perturbations because, as we have mentioned earlier 
(around (\ref{zetaS})), the entropy perturbation may lead to 
a growth of the adiabatic perturbation in time. 
In the present case, however, 
we can solve for both the adiabatic and the entropy modes, i.e. for
both $\rh$ and $\mu$ (both are constant on super-Hubble scales),
and the joint analysis of the equations
lead to $\Phi$ having an approximately scale invariant spectrum. 

\section{$SO(N)$ Symmetry} \label{SON} 
 
In this section we study another situation where 
we can completely determine the evolution of perturbations 
by solving for only two independent variables, 
without having to know each of the $N$ variables $Q_I$. 
Consider the potential term 
\be 
V = V(B), 
\ee 
where 
\be 
B \equiv \sum_I \phi_I^2, 
\ee 
so that the system of inflatons has $SO(N)$ symmetry. 
By multiplying Equation (\ref{EOMphi}) by $\phi_I$
and summing over $I$, we have 
\be 
\ddot{B}+3H \dot{B} + 4B\frac{dV(B)}{dB} + 4\dot{H} =0. \label{B} 
\ee 
Together with Equation (\ref{V}), $B$ and $H$ form a set of 
coupled differential equations.  Given $B(t_0), \dot{B}(t_0)$, and 
$H(t_0)$, we can determine $B$ and $H$ uniquely. We can then solve for 
$K$ from Equation (\ref{EOMa1}).  Hence, these equations completely
specify the dynamics of the background.  Since the dependence of $B$ on $H$ is 
via the functional form of $V$, the result is independent of the 
value of $N$. 

The one special case is $N=1$. In this case, we have 
\be 
\dot{B}= 2\phi \dot{\phi}.  
\ee 
Making use of Equation (\ref{K}), we obtain another equation for $B$ and
$H$, namely 
\be 
\dot{B}^2 = -8B\dot{H}, \label{B1} 
\ee 
which implies Equation (\ref{B}). In this case, we only need 
$B(t_0)$ and $H(t_0)$ to specify the dynamics and the phase space is 
reduced by one dimension. Equivalently, we can use 
Equations (\ref{V}) and (\ref{B}) to determine $B$ and $H$ like  in the 
case of multiple inflatons except that $\dot{B}(t_0)$ is now fixed by 
$B(t_0)$ and $H(t_0)$ through Equations (\ref{V}) and (\ref{B1}).

We emphasize that although the theory itself 
has an $SO(N)$ symmetry, the classical background is not restricted 
to be $SO(N)$ invariant. Remarkably, as we will see below, 
we can still determine the metric perturbations 
by dealing only with two variables. 
 
Let us now return to the equations for the perturbations for generic N.
Equation (\ref{EOMQ}) implies that 
\bea 
&\ddot{\rh}+9H\dot{\rh}+ 
\left( \frac{k^2}{a^2}+3\dot{H}+18H^2 
-2\frac{\dot{H}^2}{H^2} 
+\frac{\ddot{H}}{H} \right)\rh \nn \\ 
&+2\dot{\mu}+\left( 6H-2\frac{\dot{H}}{H} \right)\mu = 0, 
\label{rho} \\ 
&\ddot{\mu}+(3H-2\u)\dot{\mu}+ 
\left(\frac{k^2}{a^2}+6\dot{H}+\frac{\ddot{H}}{H} 
+\u^2-3H\u-\dot{\u}+4B\frac{d^2 V}{dB^2} \right)\mu \nn \\ 
& -4\frac{dV}{dB}\dot{\rh}-\left\{12H\frac{dV}{dB}-\frac{4}{H}B 
\left(\frac{dV}{dB}\right)^2-\frac{1}{H^2}(\dot{H}+3H^2)(\ddot{H}+6H\dot{H})\right\}\rh 
= 0. 
\label{mu} 
\eea 
Here, 
\be 
\u = \frac{d}{dt}\log(\frac{dV}{dB}) =
\frac{\frac{d^2 V}{dB^2}(\ddot{H}+6H\dot{H})}{\left(\frac{dV}{dB}\right)^2}, 
\ee 
where we have used $\frac{dV}{dB}\dot{B}=\dot{V}$. 
 
If the inverse function $B=B(V)$ exists, then
$B$, $V$, $\frac{dV}{dB}$ and $\frac{d^2 V}{dB^2}$ can be viewed as 
given functions of time, 
and thus both of the above differential equations 
can be used to solve for $\rh$ and $\mu$. In turn, these results
can then be used to determine $\zeta$, $\S$ and $\Phi$. 
Our goal is to study under which conditions the resulting spectrum
of fluctuations will be scale-invariant, starting with an initial
vacuum state on sub-Hubble scales at the initial time. One obtains
a scale-invariant spectrum if the Fourier modes of the
canonically normalized fields
representing the fluctuations undergo oscillations with constant
amplitude while on sub-Hubble scales, and are then squeezed with
amplitude proportional to the scale factor $a(t)$ on super-Hubble
scales (see e.g. \cite{MFB,RHBrev} for comprehensive reviews).
For the fields $Q_I$ these requirements imply that we want $Q_I$
to undergo damped oscillations on sub-Hubble scales (with the
amplitude decreasing as $a^{-1}(t)$, and to be squeezed as in the
case of the previous section on super-Hubble scales).
 
Assuming the slow roll conditions (\ref{slowroll}), we see $\u$ and $\dot{\u}$ 
are approximately zero and the Equations (\ref{rho}), (\ref{mu}) 
can be simplified. In terms of the variables (\ref{hat}) and conformal time, 
these equations are 
\bea 
&\hat{\rh}''+\left(k^2-2\H^2\right)\hat{\rh} 
+ 2a^4(\hat{\mu}'+2\H\hat{\mu}) = 0, 
\label{rhohat1}\\ 
&\hat{\mu}''+ 
\left(k^2-2\H^2+4a^2 B \frac{d^2 V}{dB^2} \right)\hat{\mu} - 4a^{-2} \frac{dV}{dB}\hat{\rh}'+ 
\left( \frac{4\H \frac{dV}{dB}}{a^2} + \frac{4B \left(\frac{dV}{dB}\right)^2}{\H}\right)\hat{\rh} = 0. 
\label{muhat1} 
\eea 
Notice that the primes mean derivatives with respect to $\eta$.
 
A priori, these coupled equations admit the possibility of 
having more than the two phases (sub-Hubble and super-Hubble)
described above. In the case of a single inflaton, 
there is a single transition (at Hubble radius crossing) for each
fluctuation mode 
from the oscillatory phase to the frozen phase. 
In the case of many inflatons, a fluctuation mode may experience more than 
one transition between oscillatory and frozen phases before the end of 
inflation. However, for the following explicit examples, 
we do not find multiple phases. Instead, we find that  
some fluctuation modes never experience a frozen phase, like what 
happens for a single scalar field with a very large mass. 
 
\subsection{Massive Fields} 
 
To begin, we consider the simplest case of free fields with 
\be 
V(B) = \frac{1}{2}m^2 B. 
\ee 
In this case $\frac{dV}{dB}=m^2/2$ and $\frac{d^2 V}{dB^2}=0$. 
The time evolution of $B$ is determined by Equations 
(\ref{V}) and (\ref{B}). 
Using the definition $\tilde{m} = m/H$, Equations (\ref{rhohat1}) and 
(\ref{muhat1}) can be simplified to yield 
\bea 
&\hat{\rh}''+(k^2- 2\H^2)\hat{\rh} 
+ 2a^{4}\left(\hat{\mu}'+ 2\H\hat{\mu}\right) = 0, \label{rho3} \\ 
&\hat{\mu}''+(k^2-2\H^2)\hat{\mu} 
-2\tilde{m}^2 \H^2 a^{-4}\left(\hat{\rh}'- 4\H \hat{\rh}\right) = 0. 
\label{mu3} 
\eea 
Eliminating $\hat{\mu}$ from the above two equations, we can obtain a fourth differential equation for $\hat{\rh}$: 
\bea 
&\;& \hat{\rh}''''-8\H\hat{\rh}''' 
+\left\{2k^2+4(1+\tilde{m}^2)\H^2\right\}\hat{\rh}'' 
-\left\{8k^2+8(-1+2\tilde{m}^2)\H^2 \right\}\H\hat{\rh}' \nonumber\\ 
&\;& +\left\{k^4+4k^2\H^2+8(1-2\tilde{m}^2)\H^4 \right\}\hat{\rh} = 0. 
\label{rho4} 
\eea 
Once we solve the above equation for $\hat{\rh}$, 
we can find $\hat{\mu}$ by using Equation (\ref{rho3}). 
 
To get a rough idea of the solutions to these equations, 
we consider the following limits. 
 
\bde 
\item{(1)} $\H \ll k$ (sub-Hubble), with $\tilde{m}$ fixed.  
 
Equation (\ref{rho4}) can be solved order by order in $\H/k$.
For more detailed analysis, we consider the case of exponential inflation
which is consistent with the slow roll condition. In this case 
\be \label{exp}
a= -1/(H \eta), 
\ee
and $H$ is a constant.

\bde
\item{(a)}
For $\tilde{m} \ll 1$,
it reduces to the case of a smooth potential.
We find
\bea 
&\;& \hat{\rh} \simeq c_1 \eta^{-3} \cos(k\eta+\th_1) + 
c_2 \cos(k\eta+\th_2); \label{rhoh}\\ 
&\;& \hat{\mu} \simeq 3 H^4 c_1 \cos(k\eta+\th_1). \label{muh} 
\eea 
%
Hence, both $\hat{\rh}$ and $\hat{\mu}$ are in the oscillatory phase.
The amplitude of the oscillation of $\mu$ is decreasing as $a^{-1}(t)$
as we need it to in order to obtain a scale invariant spectrum. 
(The relation between $\mu$ and $\hat{\mu}$ is given in (\ref{hat}).)
%

\item{(b)}
For $\tilde{m} \gg 1$,
the fluctuations are in the oscillatory phase
with a different amplitude evolution
\bea
&\;& \hat{\rh} \propto \eta^{-3/2} \cos(k\eta+\th_1)
\cos(\tilde{m}\log \eta+\th_2); \label{rhoh1}\\
&\;& \hat{\mu} \propto \eta^{3/2} \cos(k\eta+\th_1)
\cos(\tilde{m}\log \eta+\th_2). \label{muh1}
\eea 

\ede

\item{(2)} $\H \gg k$ (super-Hubble), with $\tilde{m}$ fixed. 
 
Under this condition, and with (\ref{exp}),
Equation (\ref{rho4}) becomes 
\bea 
&\;& \hat{\rh}''''+ \frac{8}{\eta}\hat{\rh}''' 
+\frac{4(1+\tilde{m}^2)}{\eta^2}\hat{\rh}'' 
-\frac{8(1-2\tilde{m}^2)}{\eta^3}\hat{\rh}' 
+\frac{8(1-2\tilde{m}^2)}{\eta^4}\hat{\rh} = 0. 
\eea 
Similarly, we can find the solutions of $\hat{\rh}$ and $\hat{\mu}$: 
\bea 
&\;& \hat{\rh} = c_2 \eta^{-4} 
+ c_3 \eta^{\frac{1+\sqrt{9-16\tilde{m}^2}}{2}} 
+ c_4 \eta^{\frac{1-\sqrt{9-16\tilde{m}^2}}{2}}, \\ 
&\;& \hat{\mu} = H^4\left\{c_1 \eta^{2} + 3c_2 \eta^{-1} 
+ \left(\frac{3-\sqrt{9-16\tilde{m}^2}}{4}\right) c_3 \eta^{\frac{7+\sqrt{9-16\tilde{m}^2}}{2}}\right. \nonumber \\ 
&\;&\hskip 4.2cm \left.+ \left({\frac{3+\sqrt{9-16\tilde{m}^2}}{4}}\right) c_4 \eta^{\frac{7-\sqrt{9-16\tilde{m}^2}}{2}}\right\}.  
\eea 
 
\bde 
\item{(a)} For $\tilde{m}\ll 1$, 
the evolution equations reduce to 
the slow roll scenario with a shallow potential in Sec. \ref{SP}. 
$\rh$ and $\mu$ freeze out and the wave functions are squeezed as
in single-field inflation. Hence, the resulting
spectrum is scale invariant. 
 
\item{(b)} For $\tilde{m}\gg 1$, the large mass induces an
oscillation of $\hat{\rh}$ and $\hat{\mu}$ even on these super-Hubble scales: 
\bea 
&\;& \hat{\rh} \propto \eta^{1/2} \cos(2\tilde{m} \ln\eta); \\ 
&\;& \hat{\mu} \propto \eta^{7/2} \sin(2\tilde{m} \ln\eta). 
\eea 
\ede 
\ede 
 
To summarize, in an accelerating Universe, 
a fluctuation mode with given $k$ starts 
in the oscillatory phase (quantum vacuum oscillations)
with $k \gg \H$ (Case (1)). If $\tilde{m} \ll 1$, then when
the mode crosses the Hubble radius it freezes out and its amplitude 
grows.  Thus, if $\tilde{m}  \ll 1$ at the time of Hubble radius crossing, 
the qualitative picture of the evolution of fluctuations 
is the same as in the case of a single inflaton. 
 
On the other hand, if $\tilde{m} \gg 1$ at Hubble radius crossing,  
$\rh$ keeps oscillating even on super-Hubble scales. Hence, the 
generation of classical fluctuations is suppressed. 
 
Except in the case of exponential inflation, 
generically each fluctuation mode crosses  
the Hubble radius with a different Hubble parameter $H$ 
and thus a different value of $\tilde{m}$. 
For power law inflation, i.e. $a(t) \propto t^n$ with $n>1$, 
$H$ decreases with time, and so $m \ll H$ for lower frequency modes 
which cross the Hubble radius earlier, 
and $m \gg H$ for higher frequency modes. 
Consequently, the high frequency part of the spectrum is suppressed. 
This effect has the wrong sign to explain the suppression of the 
spectrum at large wavelengths 
(observed in the recent CMB data \cite{Bennett}) compared
to what single field inflation models predict. 
 
\subsection{$\phi^4$ Theory} 
 
Finally we consider the $\phi^4$ theory, 
which is the interesting case in the context 
of the large $N$ limit mentioned in Section \ref{LargeN}. 
 
For simplicity, we ignore the mass term in the potential and write 
\be 
V(B) = \frac{\lam}{4}B^2. 
\ee 
{}From (\ref{V}) we find $B \simeq \sqrt{\frac{12}{\lam}}H$. Making use
of the definition  $\tilde{\lam} = 27\lam/H^2$, 
it follows from (\ref{rhohat1}), (\ref{muhat1}) that 
\bea 
&\hat{\rh}''+\left(k^2-2\H^2\right)\hat{\rh} 
+a^4\left\{2\hat{\mu}'+4\H\hat{\mu}\right\} = 0, \\ 
&\hat{\mu}''+\left\{k^2+(-2+\frac{4}{3}\sqrt{\tilde{\lam}})\H^2 \right\} 
\hat{\mu} 
-\frac{4}{3}\sqrt{\tilde{\lam}}\H^2 a^{-4}\left\{\hat{\rh}'- 
7\H\hat{\rh}\right\} = 0. 
\eea 
As in the previous subsection, we can combine these two equations 
to obtain a fourth order differential equation for $\hat{\rh}$:
\bea  
\hskip -2cm (k^2+ \frac{4}{3}\sqrt{\tilde{\lam}}\H^2)\hat{\rh}'''' 
&-& 8(k^2+\frac{1}{3}\sqrt{\tilde{\lam}}\H^2)\H\hat{\rh}''' 
+\left\{2k^4 + 4k^2(1+\frac{5}{3}\sqrt{\tilde{\lam}})\H^2 
+ \frac{16}{3}(4\sqrt{\tilde{\lam}}+\tilde{\lam})\H^4\right\}\hat{\rh}'' 
\nonumber\\ 
\hskip -2cm &-& \left\{ 8k^4 + 8k^2(-1+4\sqrt{\tilde{\lam}})\H^2 
+ 16(-\sqrt{\tilde{\lam}}+2\tilde{\lam})\H^4\right\}\H\hat{\rh}' 
\label{rho5} \\ 
\hskip -2cm &+& \left\{k^6 + 4k^4(1+\frac{2}{3}\sqrt{\tilde{\lam}})\H^2 
+ 8k^2(1+\frac{2}{9}\tilde{\lam})\H^4+\frac{32}{3}(-\sqrt{\tilde{\lam}} 
+2\tilde{\lam})\H^6\right\}\hat{\rh} = 0.  \nonumber
\eea 

The discussion of the growth of adiabatic perturbations 
parallels that of the previous subsection. 
The fluctuation starts in a phase of vacuum oscillations 
and then, for a certain region of wavelengths, freezes out at Hubble
radius crossing. The other wavelength modes remain in the oscillatory phase 
even after crossing the Hubble radius. 
Let us describe each phase separately. 
\bde 
\item{(1)} $k \gg \H$, with $\tilde{\lam}$ fixed. 
 
Assuming $\tilde{\lam} \ll 1$,
this is the short wavelength phase of vacuum oscillations in which 
$\hat{\rh}$ and $\hat{\mu}$ evolve in exactly the same 
as given in the previous subsection (Equations (\ref{rhoh}) and (\ref{muh})). 
This is far from surprising since only terms with leading $k$ dependence 
survive in this limit. 
If $\tilde{\lam} \gg 1$,
the result is the same as (\ref{rhoh1}) and (\ref{muh1})
with $\tilde{m}^2$ replaced by $\frac{2}{3}\sqrt{\tilde{\lam}}$.

\item{(2)} $k \ll \H$, with $\tilde{\lam}$ fixed. 
 
To be specific, we again consider the case of exponential inflation.
In this region (long wavelength region), Equation (\ref{rho5}) becomes 
\bea 
&\;& \hskip -2cm \hat{\rh}''''+ \frac{10}{\eta}\hat{\rh}''' 
+\frac{4(4+\sqrt{\tilde{\lam}})}{\eta^2}\hat{\rh}'' 
+\frac{12(-1+2\sqrt{\tilde{\lam}})}{\eta^3}\hat{\rh}' 
+\frac{8(-1 + 2\sqrt{\tilde{\lam}})}{\eta^4}\hat{\rh} = 0. 
\eea 
As in the previous subsection, we can find the 
solutions of $\hat{\rh}$ and $\hat{\mu}$: 
\bea 
&\;& \hat{\rh} = c_1 \eta^{-1} +c_2 \eta^{-4} 
+ c_3 \eta^{\frac{1+\sqrt{9-16\sqrt{\tilde{\lam}}}}{2}} 
+ c_4 \eta^{\frac{1-\sqrt{9-16\sqrt{\tilde{\lam}}}}{2}}, \\ 
&\;& \hat{\mu} = H^4\left\{6c_1 \eta^{2} + 3c_2 \eta^{-1} 
+ \left(\frac{3-\sqrt{9-16\sqrt{\tilde{\lam}}}}{4}\right) c_3 \eta^{\frac{7+\sqrt{9-16\sqrt{\tilde{\lam}}}}{2}}\right. \nonumber \\ 
&\;&\hskip 4.2cm \left.+ \left({\frac{3+\sqrt{9-16\sqrt{\tilde{\lam}}}}{4}}\right) c_4 \eta^{\frac{7-\sqrt{9-16\sqrt{\tilde{\lam}}}}{2}}\right\},  
\eea 
 
Again there are two cases: 
 
\bde 
\item{(a)} $\tilde{\lam} \ll 1$. 
In this case, the fluctuations are frozen and their amplitude grows 
(with the leading  time dependence the same, but the sub-leading terms
slightly different from what occurs in the 
slow roll scenario for a flat potential in Sec. \ref{SP}). 
The spectrum is still scale invariant. 
 
\item{(b)} $\tilde{\lam} \gg 1$. 
In this case, the fluctuations are oscillatory even on super-Hubble
scales, with 
\bea 
&\;& \hat{\rh} \propto \eta^{1/2} \cos(2\tilde{\lam}^{1/4} \ln\eta); \\ 
&\;& \hat{\mu} \propto \eta^{7/2} \sin(2\tilde{\lam}^{1/4} \ln\eta). 
\eea 
\ede 
\ede 
This result is qualitatively similar to what was obtained in 
the previous subsection. 
 
\section{Discussion} 
 
In this paper we have studied inflationary dynamics 
of a system of $N$ scalar fields coupled to Einstein gravity 
as a model for the very early Universe. In the large $N$ limit,
the probability for slow roll inflation increases. 
We showed that if the scalar fields respect $SO(N)$ symmetry, 
the phase space for the homogeneous, isotropic background 
is three dimensional ($B, \dot{B}, H$), 
only one dimension more than the case of a single scalar field. 
Furthermore, for arbitrary $N$, 
we showed that only two independent variables ($\rh, \mu$) 
are needed to determine both adiabatic and entropy perturbations,
and not $N$ as could be naively expected. 
In the case of a single inflaton field, only one variable is needed
since there is (on super-Hubble scales) only the adiabatic fluctuation
mode. This simplification allows us to analyze 
the fluctuation spectrum and 
find its deviation from the case of a single inflaton. 
Note that this model is independent of $N$ 
in terms of the appropriate variables. 
It will be interesting to build models of this class 
to explain experimental data and compare with single-inflaton models. 
 
Note that, as shown in Section \ref{LargeN}, initial conditions with
field values of order one in Planck units
may induce slow roll inflation in the large $N$ limit. 
It would be interesting to consider the possibility of inflationary models 
with trivial classical background, 
and where inflation is induced purely by quantum fluctuations. In
this case, a new formulation for the generation of fluctuations has to be 
constructed (the usual theory requires a non-vanishing classical background
matter field). We leave this interesting question for future study
(for early work on this issue see \cite{Mary}). 

Upon completion of this work, we were informed of a paper \cite{Anupam}
which deals with complementary aspects of large N inflation.
 
\section*{Acknowledgment} 
 
The authors thank Chong-Sun Chu and Steven Hsu 
for helpful discussions. 
The work of PMH and HCK is supported in part by 
the National Science Council 
and the National Center for Theoretical Sciences. 
Taiwan, R.O.C. 
The work of PMH is also partly supported by 
the Center for Theoretical Physics 
at National Taiwan University, 
the CosPA project of the Ministry of Education, Taiwan, 
and the Wu Ta-Yu Memorial Award of NSC. R.B. is supported
in part by the U.S. Department of Energy under
Contract DE-FG02-91ER40688, TASK A

\end{document}